UNIVERSITY
of
GLASGOW

Department of Physics & Astronomy
Experimental Particle Physics Group
Kelvin Building, University of Glasgow,
Glasgow, G12 8QQ, Scotland
Telephone: +44 (0)141 339 8855 Fax: +44 (0)141 334 9029

# MEASUREMENT OF THE PROTON STRUCTURE FUNCTION $F_2$ AND THE EXTRACTION OF THE GLUON DENSITY AT SMALL $x$

ZEUS Collaboration
talk given at Photon'95, 8–13th April 95, by J. I. FLECK
*Department of Physics and Astronomy,*
*University of Glasgow, Glasgow, UK.*
E-mail: fleck@desy.de

## Abstract

In the following article we describe our measurement of the proton structure function $F_2$ and of the gluon momentum density $xg$ in $ep$ collisions with the ZEUS detector at HERA in 1993. The results for $F_2$ confirm our measurement from the previous year but with much higher statistics and show a strong rise towards small $x$. The gluon momentum density $xg$ is measured for the first time in a range from $4 \cdot 10^{-4} < x < 10^{-2}$ at a value of $Q^2 = 20$ GeV$^2$.

# 1 Introduction

In 1993, electrons of 26.7 GeV and protons of 820 GeV were brought into collision at HERA. 84 paired bunches out of 210 possible were filled and in addition 10 electron and 6 proton bunches were left unpaired for background studies.

The measurements described here were made using the ZEUS detector, which is described in detail in [1] and [2]. The main components used for our analyses were the vertex detector (VXD), the central tracking detector (CTD), the uranium calorimeter (CAL), a lead scintillator beam monitor (C5) and the luminosity detector.

With the data taken in 1993 a measurement of the proton structure function $F_2$ was possible with high precision in a region of $7 < Q^2 < 10^4$ GeV$^2$ and $x$ values as low as $3 \times 10^{-4}$. The total integrated luminosity in 1993 was 0.54 pb$^{-1}$.

# 2 Kinematics

The processes in deep inelastic scattering (DIS) are described by the following kinematic variables.

$$Q^2 = -q^2 = -(k-k')^2, \quad x = \frac{Q^2}{2P \cdot q}, \quad y = \frac{P \cdot q}{P \cdot k} \tag{1}$$

$k$ and $k'$ denote the four-momenta of the incoming and the final state lepton, $P$ is the four-momentum of the incoming proton and $y$ is the fractional energy transfer in the proton rest frame.

As the ZEUS detector is a nearly hermetic detector (99.7 %), there exist various methods to reconstruct the kinematic variables. For our analysis we have used two different methods:

1. electron method; (in this method only the scattered electron is observed and its scattering angle and energy are measured.)
2. double angle (DA) method; (this relies on the measurement of two scattering angles, that of the scattered electron and that for the final hadronic state. In the naive Quark Parton Model the latter is the angle of the scattered quark.)

Method 1 gives a better resolution in $x$ at low $Q^2$ while method 2 is less sensitive to the absolute calorimeter energy scale and gives better resolution over the whole $x - Q^2$ plane. Comparing the results obtained by both methods gives a good systematic check of our analysis.

The range of $x$ and $Q^2$ that is accessible with these two methods is shown in figure 1.

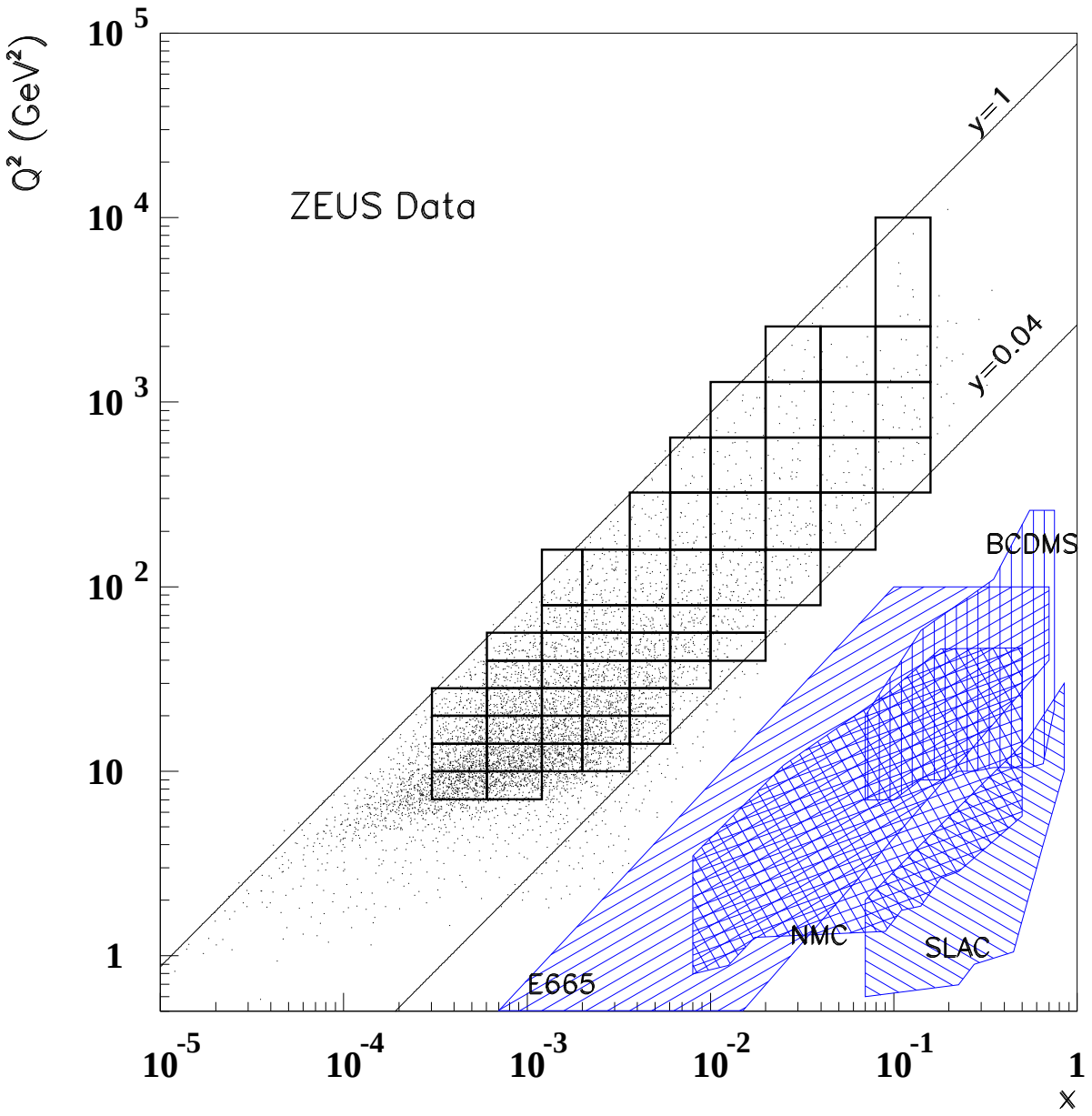

Figure 1: Distribution of events in the $x$ and $Q^2$ plane. Also shown are the bins used and the range covered by previous experiments.

## 3 Event Selection

The event selection is dominated by two selection steps, the trigger and the offline selection. The ZEUS trigger is a three step algorithm and has an efficiency greater than 95% for DIS events in the $x$ and $Q^2$ region discussed here. A total of $2 \cdot 10^6$ events passed the DIS trigger. To obtain an event sample with little background and high purity, several cuts were applied. The most important are listed below:

- $35 < \delta < 60$ GeV, $\delta = \sum_i E_i(1 - \cos\theta_i)$, where $E_i$ denotes the energy deposit in calorimeter cell $i$ and $\theta_i$ the polar angle of the centre of the cell with respect to the event vertex; (For fully contained DIS events $\delta \approx 2E_e = 53.4$ GeV, whereas photoproduction events peak at low values of $\delta$, as the scattered electron remains within the rear beampipe.)
- $E'_e > 5$ GeV; the energy of the scattered electron should be greater than 5 GeV for a reliable electron finding and a low background from photoproduction;
- $y_{JB} > 0.04$ to ensure a good measurement of the hadronic angle and of $x$.

A total of 46k DIS NC events corresponding to an integrated luminosity of 0.54 pb$^{-1}$ survived all cuts.

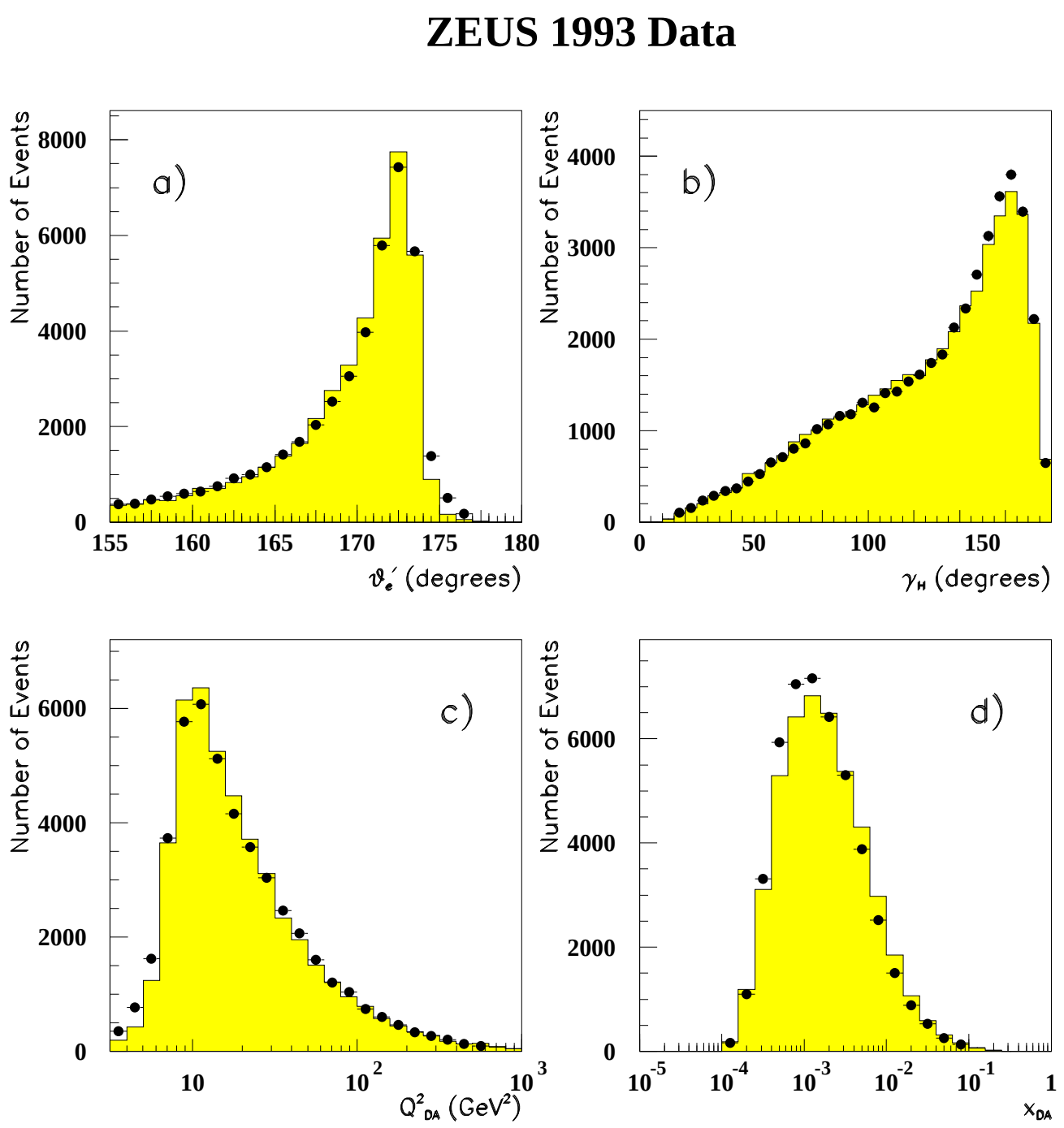

Figure 2: Distribution of the primary measured quantities, a) angle of the scattered electron $\vartheta'_e$, b) angle of the hadronic final state $\gamma_H$, c) $Q^2$, and d) $x$. The dots are the data and the shaded areas represent the MC.

The distributions for the measured and the reconstructed variables for the DA method for data and MC are shown in figure 2. The good agreement between the data and MC distributions demonstrates our good understanding of the relation between MC and data. The structure function used as an input for the MC distribution has been reweighted to the measured structure function.

## 4 Background

The main source of background originates from photoproduction events. Figure 3 shows the distribution of DIS and photoproduction events in a given $x$ and $Q^2$ bin after all cuts, except for the cut on $\delta$, have been applied. To calculate the number of photoproduction events with $\delta > 35$ GeV a fit is made to the data, shown as black dots, in each bin of the $x - Q^2$ plane. From the total number of events in each bin, the number of photoproduction events as calculated from the fit, is subtracted. The photoproduction background can rise to 12% at high $y$, but is a few percent for most of the bins. Other sources of background, like cosmic ray events and halo muons, contribute much less and are efficiently removed.

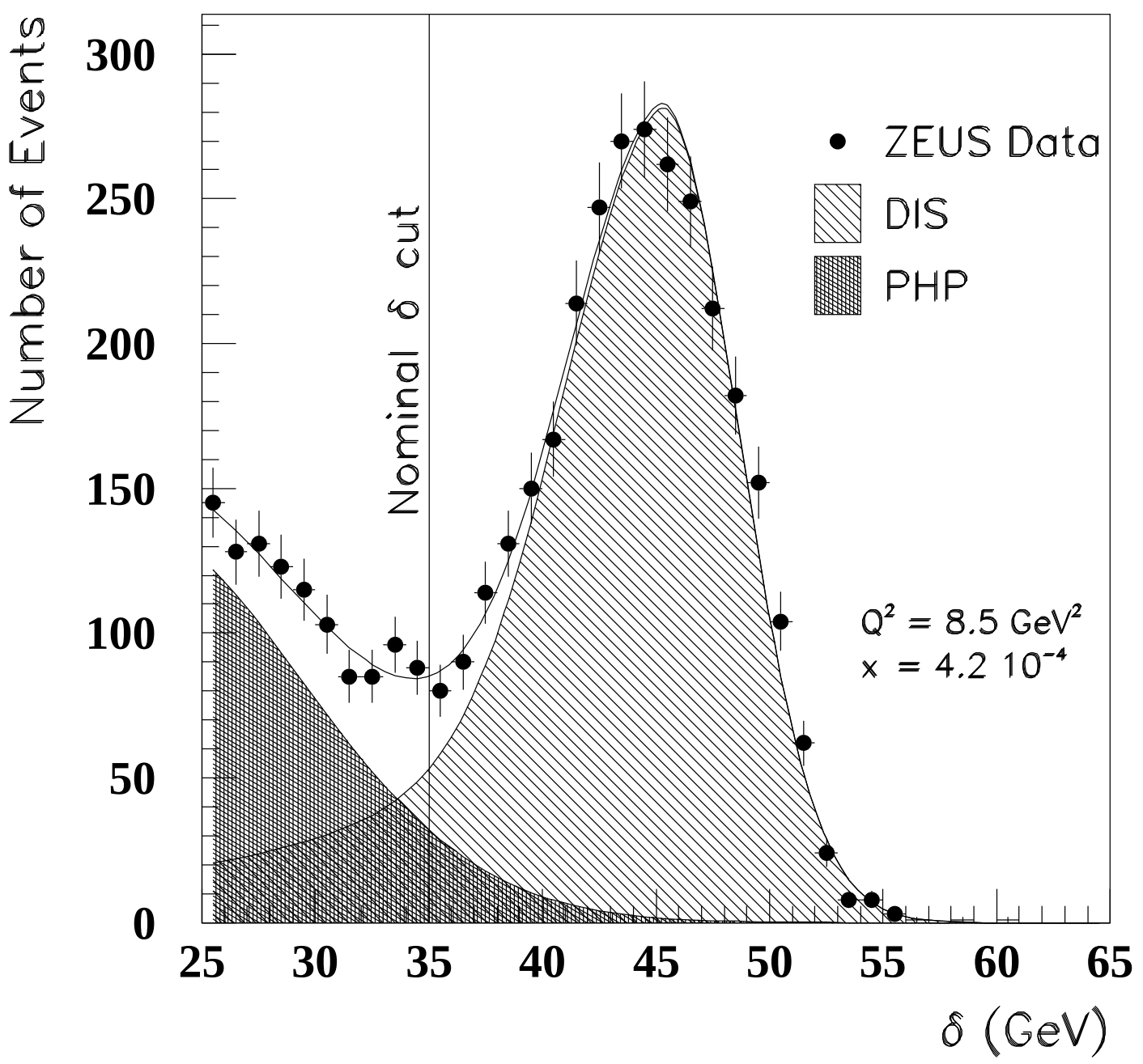

Figure 3: Distribution of DIS and photoproduction (php) events in a given $x - Q^2$ bin. The black dots show the data and the solid line represents the fit that is applied to the data.

## 5 The Proton Structure Function $F_2$

The proton structure function $F_2$ is related to the double differential cross section for inclusive $ep$ scattering including photon and $Z^0$ exchange via the following formula:

$$\frac{d^2\sigma}{dx\,dQ^2} = \frac{2\pi\alpha^2}{xQ^4}\left[Y_+\mathcal{F}_2(x,Q^2) - y^2\mathcal{F}_L(x,Q^2) + Y_- x\mathcal{F}_3(x,Q^2)\right](1+\delta_r(x,Q^2)) \qquad (2)$$

with $Y_\pm = 1 \pm (1-y^2)$. $\mathcal{F}_L$ is the longitudinal structure function, $\mathcal{F}_3$ is the parity violating term, and $\delta_r$ is the radiative correction. For comparison with other experiments we measure $F_2$ containing only contributions from virtual photon exchange. Therefore we have to absorb the contributions from the $Z^0$ exchange in a correction factor $\delta_Z$. This correction is 6% for the highest $Q^2$ bin shown in this analysis and lower for all the other bins. Also the corrections due to the longitudinal structure function and the radiative corrections are small (lower than 15% and 5% respectively), and can be absorbed in correction factors $\delta_{F_L}$, from QCD prescriptions, and $\delta_r$, from the HERACLES Monte Carlo. This results in the following equation for $F_2$:

$$\frac{d^2\sigma}{dx\,dQ^2} = \frac{2\pi\alpha^2 Y_+}{xQ^4}F_2(x,Q^2)(1-\delta_{F_L}+\delta_Z)(1+\delta_r) \qquad (3)$$

To derive $F_2$ from the observed double differential cross section, an unfolding method is used.

The results for the DA method are shown in figure 4. Our most important systematic check is the comparison of the values of $F_2$ for the DA and the electron method. These two methods use overlapping data samples, but the reconstruction of the kinematic variables is largely independent. Within the statistical errors there is very good agreement between the two methods.

## 6 Gluon Momentum Density

In zeroth order, $F_2$ is constant in $Q^2$ and the gluons do not contribute to the cross section. The scaling violations, as shown in figure 5, are a higher order process (called leading order,

ZEUS 1993 $F_2$

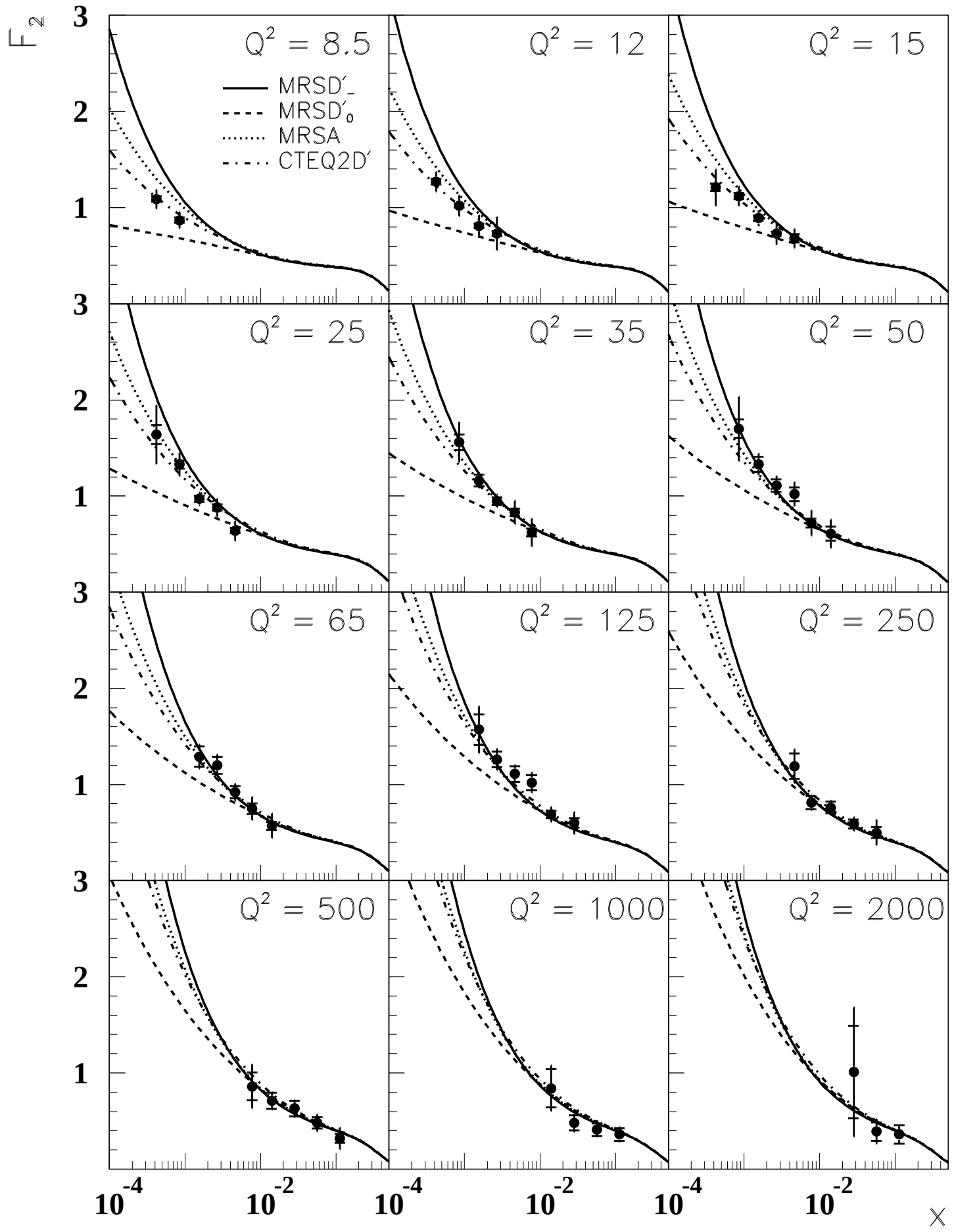

Figure 4: Proton structure function $F_2$ as a function of $x$. The inner error bars show the statistical error only, while the outer bars show the statistical and the systematic error added in quadrature. $Q^2$ is given in $\mathrm{GeV}^2$.

LO, for our purpose). At small values of $x$, $x < 10^{-2}$, they are dominated by quark pair production from gluons leading to positive scaling violations with increasing $Q^2$. Therefore the gluon momentum density, $xg(x, Q^2)$, can directly be extracted from the slope $dF_2/d\ln Q^2$.

We compare different methods, the approximate method of Prytz [5], EKL [6], and a fit making use of the full NLO GLAP [7] evolution equation.

The method of Prytz uses the approximation that at low values of $x$ the GLAP evolution equation is dominated by the gluonic terms. Consequently the quark splitting functions are neglected and only the gluon splitting functions are considered. Using a LO result for $P_{qg}$, the gluon splitting function and making a Taylor expansion of the gluon momentum density $zg(z, Q^2)$ at $z = x/2$ results in the following equation in LO:

$$xg(x, Q^2) \approx \frac{dF_2(x/2, Q^2)/d\ln Q^2}{(40/27)\alpha_s/4\pi}, \tag{4}$$

where the gluon density at $x$ is calculated from the slope of $F_2$ at $x/2$.

Equation 4 has been extended to NLO, still neglecting quark contributions. In the NLO GLAP fit, all the $F_2$ data from ZEUS taken in 1993 have been used. To make a constraint at large values of $x$, NMC data [8] on $F_2$ were included. The fit of the parton momentum distributions as a function of $x$ has been performed at a value of $Q_0^2 = 7$ GeV$^2$ assuming a functional form of the parton distribution as used by many parametrisations [4].

The power of the rise of the gluon momentum density with decreasing $x$ has been calculated

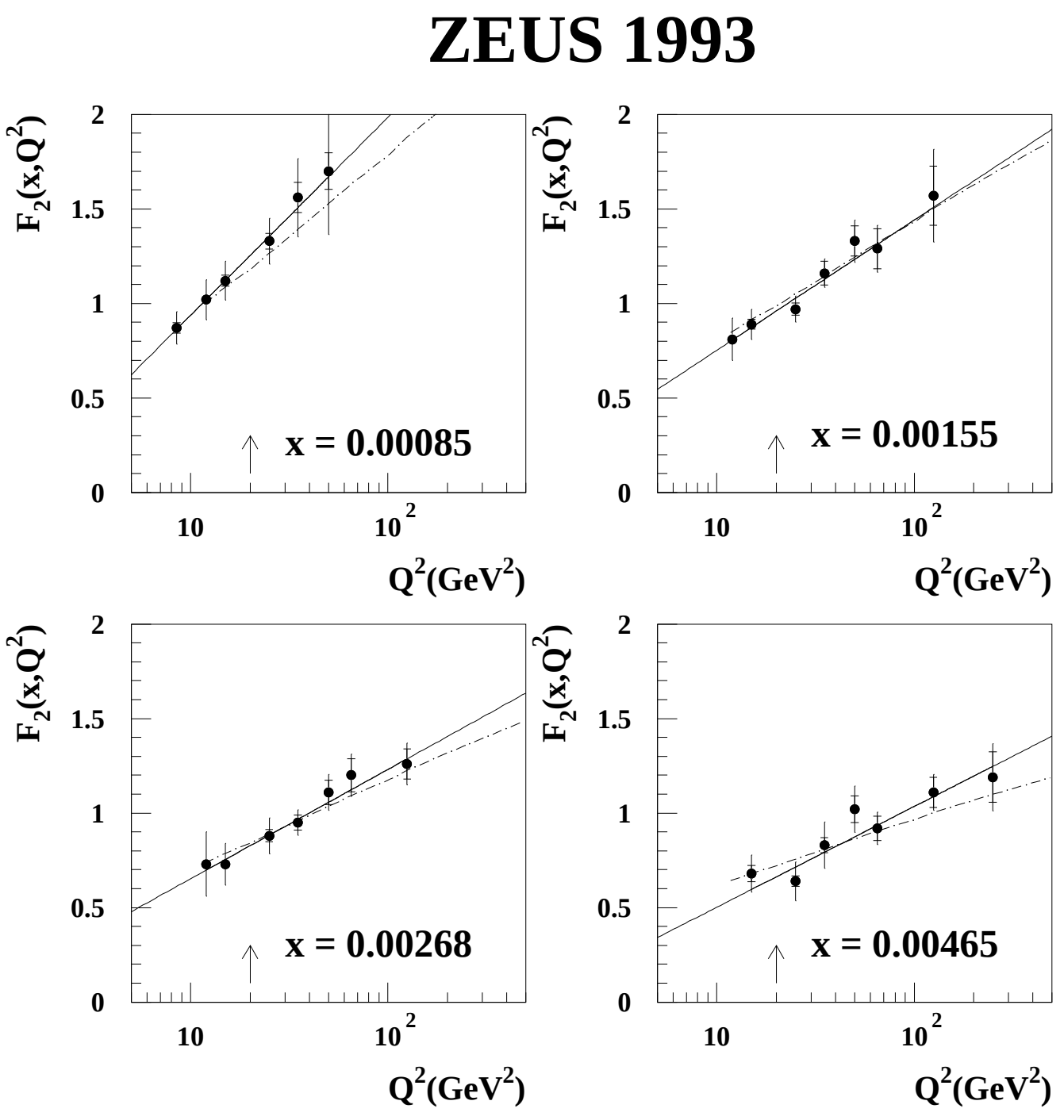

Figure 5: Scaling violation of the proton structure function $F_2$ as a function of $Q^2$ for fixed values of $x$ as given in the figure. The solid lines show a straight line fit, while the dashed lines show the result of the global NLO fit. The arrows indicate $Q^2 = 20$ GeV$^2$.

and found to be $xg \sim A \cdot x^{-0.35^{+0.10}_{-0.04}}$.

The results from the global fit, from the method of Prytz and from that of EKL are shown in figure 6 for $Q^2 = 20$ GeV$^2$. The shaded band shows the error on the global fit. The inner error bars on the points show the statistical error only, while the outer error bars show the statistical and the systematic error added in quadrature. A good agreement between all methods is observed.

## 7 Conclusions

With the 1993 ZEUS data the strong rise of $F_2$ with decreasing $x$ already observed in the previous year has been confirmed and is clearly visible over a large range of $Q^2$. The measured values of $F_2$ are in agreement with the parton distribution parametrisations from GRV, MRSA and CTEQ2D$'$. The observed scaling violations have made it possible to derive the gluon momentum density in a new regime of $x$ at $Q^2$ of 20 GeV$^2$. $xg(x, Q^2)$ shows a steep rise at low $x$ when compared to the measurements from NMC at large $x$.

## References

[1] *The ZEUS Detector, Status Report 1993*, DESY 1993.

[2] ZEUS Collaboration, M. Derrick et al., *Phys. Lett.* **B 293** (1992) 465.

[3] ZEUS Collaboration, M. Derrick et al., *Z. Phys.* **C 65** (1995) 379.

[4] ZEUS Collaboration, M. Derrick et al., *Phys. Lett.* **B 345** (1995) 576.

[5] K. Prytz, *Phys. Lett.* **B311** (1993) 286 and **B332** (1994) 393.

[6] R.K. Ellis, Z. Kunszt and E.M. Levin, *Nucl. Phys.* **B420** (1994) 517.

[7] V.N. Gribov and L.N. Lipatov, *Sov. J. Nucl. Phys.* **15** (1972) 438, 675;
L.N. Lipatov, *Sov. J. Nucl. Phys.* **20** (1975) 94;
G. Altarelli and G. Parisi, *Nucl. Phys.* **B126** (1977) 298.

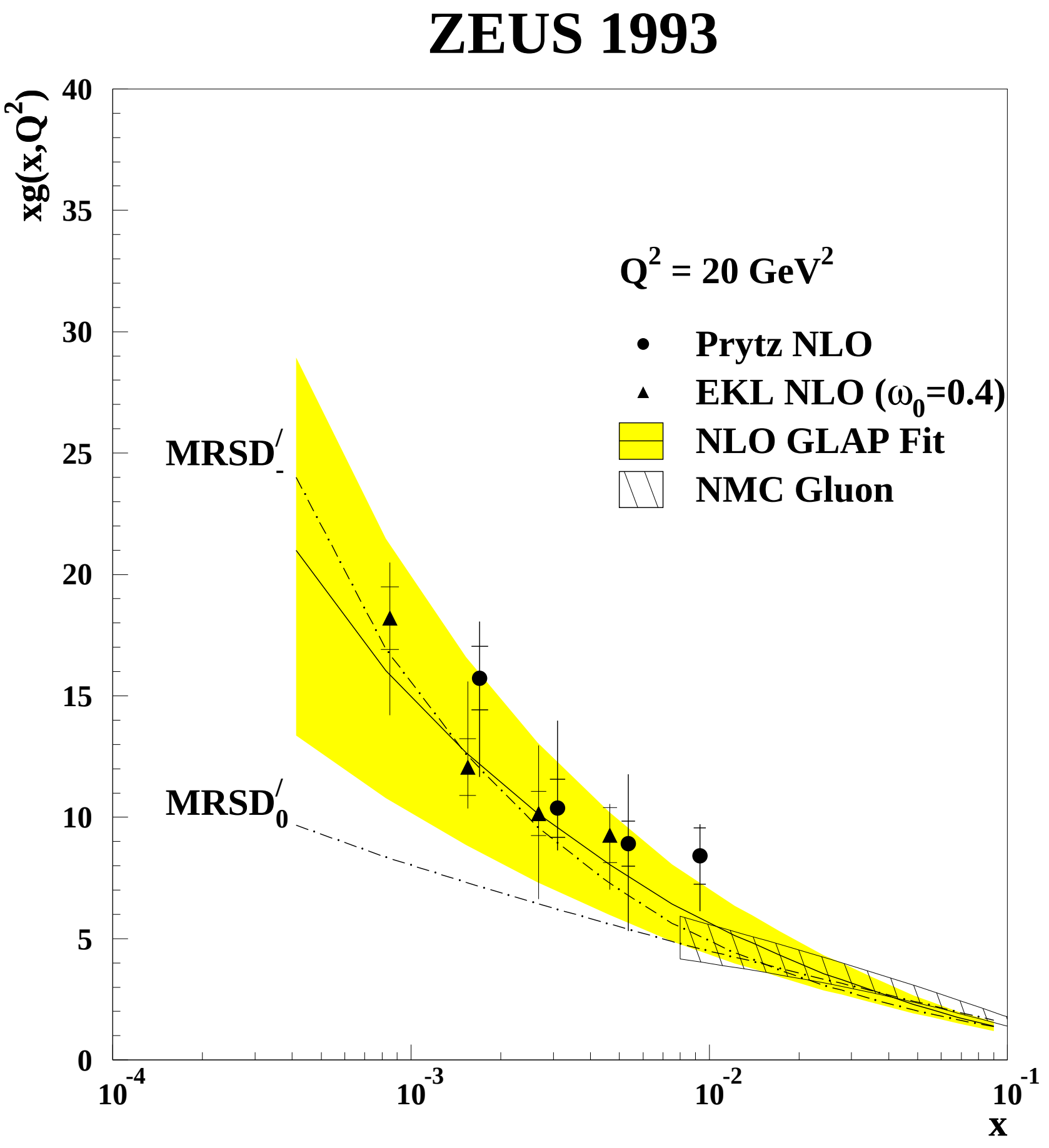

Figure 6: Gluon momentum density as a function of $x$ at $Q^2 = 20\ \mathrm{GeV}^2$

[8] NMC Collaboration, P. Amaudruz et al., *Phys. Lett.* **B 295** (1992) 159 and preprint CERN PPE/92-124.